\documentclass[rnote,traditabstract]{aa}  
\usepackage{graphicx}
\usepackage[latin1]{inputenc}
\usepackage{amsmath}
\usepackage{amsfonts}
\usepackage{amssymb}
\usepackage{natbib}
\usepackage{txfonts}
\usepackage{tabularx}
\usepackage{subfig}

\newcommand{\parcsec}{\mbox{$\stackrel{\prime\prime}{\textstyle .}$}}

\begin{document}
\title{EMCCD photometry reveals two new variable stars in the crowded central region of the globular cluster NGC 6981\thanks{Based on data collected by MiNDSTEp with the Danish 1.54 m telescope} }
   \author{Jesper~Skottfelt  \inst{1,2} \thanks{email: skottfelt@astro.ku.dk}
   \and
          D.~M.~Bramich \inst{3} \thanks{email: dbramich@eso.org}
          \and
          R.~Figuera~Jaimes \inst{3,4} \thanks{email: rfiguera@eso.org; robertofiguera@gmail.com}
          \and
 	  U.~G.~J\o{}rgensen \inst{1,2}  \thanks{email: uffegj@nbi.dk}
 	  \and
 	  N.~Kains \inst{3}
 	  \and
 	  K.~B.~W~Harps\o{}e \inst{1,2}
	  \and C.~Liebig \inst{4}
 	  \and M.~T.~Penny \inst{5}	  
	  \and K.~A.~Alsubai \inst{6}
	  \and J.~M.~Andersen \inst{7,2}
	  \and V.~Bozza \inst{8,9}
	  \and P.~Browne \inst{4}
	  \and S.~Calchi~Novati \inst{10}
	  \and Y.~Damerdji \inst{11}
	  \and C.~Diehl \inst{12,13}	  
	  \and M.~Dominik \inst{4} \thanks{Royal Society University Research Fellow}
	  \and A.~Elyiv \inst{11,14}
	  \and E.~Giannini \inst{12}
	  \and F.~Hessman \inst{15}
	  \and T.~C.~Hinse \inst{16}
	  \and M.~Hundertmark \inst{4}
	  \and D.~Juncher \inst{1,2}
	  \and E.~Kerins \inst{17}
	  \and H.~Korhonen \inst{1,2}
	  \and L.~Mancini \inst{18}
	  \and R.~Martin \inst{19}
	  \and M.~Rabus \inst{20}
	  \and S.~Rahvar \inst{21}
	  \and G.~Scarpetta \inst{10,8,9}
	  \and J.~Southworth \inst{22}
	  \and C.~Snodgrass \inst{23}
	  \and R.~A.~Street \inst{24}
	  \and J.~Surdej \inst{11}
	  \and J.~Tregloan-Reed \inst{22}
	  \and C.~Vilela \inst{22}
	  \and A.~Williams \inst{19}
    }

   \institute{Niels Bohr Institute, University of Copenhagen, Juliane Maries Vej 30, 2100 K\o{}benhavn \O{}, Denmark 
	      \and
             Centre for Star and Planet Formation, Natural History Museum, University of Copenhagen, \O{}stervoldgade 5-7, 1350 K\o{}benhavn~K, Denmark 
              \and
             European Southern Observatory, Karl-Schwarzschild-Stra\ss{}e 2, 85748 Garching bei M\"{u}nchen, Germany 
              \and 
             SUPA, University of St Andrews, School of Physics \& Astronomy, North Haugh, St Andrews, KY16 9SS, United Kingdom 
             \and Department of Astronomy, Ohio State University, 140 W. 18th Ave., Columbus, OH 43210, USA 
             \and Qatar Foundation, PO Box 5825, Doha, Qatar 
             \and Department of Astronomy, Boston University, 725 Commonwealth Avenue, Boston, MA 02215, USA 
             \and Dipartimento di Fisica ''E. R. Caianiello'', Universit\`a di Salerno, Via Ponte Don Melillo, 84084-Fisciano (SA), Italy 
             \and Istituto Nazionale di Fisica Nucleare, Sezione di Napoli, Napoli, Italy 
             \and Istituto Internazionale per gli Alti Studi Scientifici (IIASS), 84019 Vietri Sul Mare (SA), Italy 
             \and Institut d'Astrophysique et de G\'eophysique, Universit\'e de Li\`ege, All\'ee du 6 Ao\^ut, B\^at. B5c, 4000 Li\`ege, Belgium 
             \and Astronomisches Rechen-Institut, Zentrum f\"ur Astronomie der Universit\"at Heidelberg, M\"onchhofstr. 12-14, 69120 Heidelberg, Germany 
             \and Hamburger Sternwarte, Universit\"at Hamburg, Gojenbergsweg 112, 21029 Hamburg, Germany 
             \and Main Astronomical Observatory, Academy of Sciences of Ukraine, vul. Akademika Zabolotnoho 27, 03680 Kyiv, Ukraine 
             \and Institut f\"ur Astrophysik, Georg-August-Universit\"at G\"ottingen, Friedrich-Hund-Platz 1, 37077 G\"ottingen, Germany 
             \and Korea Astronomy and Space Science Institute, Daejeon 305-348, Republic of Korea 
             \and Jodrell Bank Centre for Astrophysics, University of Manchester, Oxford Road, Manchester M13 9PL, UK 
             \and Max Planck Institute for Astronomy, K\"onigstuhl 17, 69117 Heidelberg, Germany 
             \and Perth Observatory, Walnut Road, Bickley, Perth 6076, WA, Australia 
             \and Departamento de Astronom\'ia y Astrof\'isica, Pontificia Universidad Cat\'olica de Chile, Av. Vicu\~na Mackenna 4860, 7820436 Macul, Santiago, Chile 
             \and Department of Physics, Sharif University of Technology, P. O. Box 11155-9161 Tehran, Iran 
             \and Astrophysics Group, Keele University, Staffordshire, ST5 5BG, UK 
             \and Max-Planck-Institute for Solar System Research, Max-Planck Str. 2, 37191 Katlenburg-Lindau, Germany 
             \and Las Cumbres Observatory Global Telescope Network, 6740 Cortona Drive, Suite 102, Goleta, CA 93117, USA 
        }
        
   \date{Received 7 March 2013 /
	  Accepted 12 April 2013 / 
	  Erratum 2 Sep 2013}

   \authorrunning{J.Skottfelt et al.}
   \titlerunning{EMCCD photometry reveals new RR Lyrae stars in NGC~6981}
   
   \abstract{Two previously unknown variable stars in the crowded central region of the globular cluster NGC~6981 are presented.
             The observations were made using the Electron Multiplying CCD (EMCCD) camera at the Danish 1.54m Telescope at La Silla, Chile.
             The two variables were not previously detected by conventional CCD imaging because of their 
             proximity to a bright star. This discovery demonstrates that EMCCDs are a powerful tool for performing high-precision time-series photometry in crowded
             fields and near bright stars, especially when combined with difference image analysis (DIA).}


 \maketitle

\section{Introduction}

A census of the variable stars in the globular cluster NGC~6981 was performed by \citet{Bramich2011} (hereafter B11).
Using data from 10 nights of observations with a conventional CCD they were able to confirm the variability of 29 stars and refute
the suspected variability of 20 others. Furthermore, 11 new RR Lyrae stars and 3 new SX Phoenicis stars were found in the study,
bringing the total number of confirmed variable stars in NGC~6981 to 43. 

A problem with using a conventional CCD is that to obtain a reasonable signal-to-noise ratio ($S/N$) for the fainter objects in an image, the
pixels in the brightest stars may well be saturated. For DIA, which is currently the best way to extract precise photometry in crowded star fields
(e.g. towards the Galactic bulge, in the central regions of globular clusters, etc.), the saturation of the brightest stars is even more
problematic because the saturated pixels affect nearby pixels during the convolution of the reference image. Hence we cannot perform
photometric measurements using DIA near saturated stars in conventional CCD images, which has a negative impact on the completeness
of variability studies in crowded fields. 
In B11, there are 4 saturated stars in the central region of their $V$ reference image for NGC~6981, and it is therefore conceivable that their variable star census could have missed relatively bright variable stars (e.g. RR Lyraes).
An improved completeness in variability studies makes it possible to draw firmer conclusions about Oosterhoff classification \citep{Smith1995} and to 
examine whether there is a gradient in the physical properties between the central and outer parts of the cluster.

Electron-Multiplying CCDs (EMCCDs) are conventional CCDs with an extended serial register where the signal is amplified by impact
ionisation before it is read out. This means that the readout noise is negligible when compared to the signal, even at very high 
readout speeds (10-100 frames/s), enabling the possibility of high frame-rate imaging.
Numerous articles have described the possibility of using EMCCDs to obtain very high spatial resolution; see for instance \citet{Mackay2004, Law2006A}.
However, using EMCCDs to perform precise time-series photometry without throwing away photons (i.e. {\it not} Lucky Imaging) is
a new area of investigation and the applications are just starting to be explored.

With high frame-rate imaging much brighter stars can be observed without saturating the CCD and the individual exposures can be combined into stacked images at a later
stage in order to achieve the required $S/N$ for the objects of interest. We note that EMCCD exposures need to be calibrated in a different way to
conventional CCD imaging data. The algorithms required to do this have already been developed and described by \citet{Harpsoe2012}.

A previous attempt to study variability in the central region of a globular cluster using EMCCD data has been made by \citet{Diaz2012}. They used FastCam at the 2.5m Nordic Optical Telescope to obtain 200000 exposures of the globular cluster M15 with an exposure time of 30~ms.
To study the variable stars, they made a Lucky Imaging selection of the 7\%
sharpest images in each time interval of 8.1 minutes. This resulted in 20
combined images each of exposure time 21~s, where each image comes from the
combination of 700 short exposures. To extract the photometry they used standard
DAOPHOT PSF fitting routines. They did not find any new variable stars and no analysis
of the photometric precision achieved is offered.

The DIA technique, first introduced by \citet{Alard1998}, has been improved by revisions to the algorithm presented by \citet{Bramich2008,Bramich2012} and is the optimal way to perform photometry with EMCCD data in crowded fields.
This method uses a numerical kernel model instead of modelling the kernel as a combination of Gaussian basis functions and can thus give better photometric
precision even in very crowded regions \citep{Albrow2009}. The method is also especially adept at modelling images with PSFs that are not well approximated by a Gaussian.

Using the superior resolution provided by high frame-rate imaging EMCCDs in tandem with DIA we can probe the surroundings of bright stars for
variable stars which are inaccessible with conventional CCD imaging. Using this technique we are able to present EMCCD photometry of two new RR Lyrae
stars in the central region of NGC~6981.

\section{Data and Reductions}

The data were obtained over two half nights (26th and 27th August 2012) at the Danish 1.54m Telescope at La Silla Observatory, Chile, 
using the Andor Technology iXon+ model 897 EMCCD camera. The imaging area of the camera is 512x512 16$\mu m$ pixels with a pixel scale
of $0\parcsec09$ which gives a $45 \times 45$ arcsec$^2$ field-of-view (FOV). With such a small FOV, we chose to target the crowded central region of NGC~6981 including the
saturated stars from B11. The camera is equipped with a special long-pass filter with a cut-on wavelength of 650nm. The cut-off wavelength is determined by the sensitivity of the camera which drops to zero 0\% at 1050 nm over about 250 nm. 
The filter thus corresponds roughly to a combination of the SDSS i'+z' filters \citep{Bessell2005}.
A total of 44 observations with a frame-rate of 10 Hz were obtained. Each observation contains between 3000 to 3500 exposures. 

Using the algorithms described in \citet{Harpsoe2012}, each exposure is bias, flat, and tip-tilt corrected, and the instantaneous image quality (PSF width) is found.
Then, for each observation, the exposures are combined into images in two distinct ways:
\begin{description}
\item{Quality-binned: exposures are grouped according to a binning in image quality and combined to produce images that cover a range in point-spread-function (PSF) width.}
\item{Time-binned: exposures are grouped into time bins of width 2 minutes to achieve a reasonable $S/N$ at the brightness of the RR Lyrae stars. As opposed to Lucky Imaging, all frames are used. This gives a total of 125 data points in each light curve. }
\end{description}

To extract the photometry from the time-binned images we used the {\tt DanDIA} pipeline\footnote{{\tt DanDIA} is built from the DanIDL library of IDL routines available at http://www.danidl.co.uk}
\citep{Bramich2008,Bramich2012}. The pipeline has been modified to stack the sharpest of the quality-binned images to create a high-resolution 
reference image from which the reference fluxes and positions of the stars are measured. 
The reference image, convolved with the kernel solution, is subtracted from each of the time-binned images to create difference images,
and, in each difference image, the differential flux for each star is measured by scaling the PSF at the position of the star (see B11 for details).
Note that we have further modified the {\tt DanDIA} software to employ the appropriate noise model for EMCCD data \citep{Harpsoe2012}.

\section{Results} \label{sec:results}

\begin{table}
 \caption{Details of the two new variable stars found in NGC~6981. 
 \label{table:details}}
 \begin{tabularx}{\linewidth}{l l l l l l l l}
  \hline \\[-7pt]
  Variable & Var.               & RA          & Dec.        & $T_{\mbox{\scriptsize max}}$ & P      \\
  Star ID  & Type                   & (J2000.0)   & (J2000.0)   & (d)                          & (d)    \\[1pt]
  \hline \\[-7pt]
  V57      & RR1                    & 20 53 27.38 & -12 32 13.3 & 6166.779                  & 0.334  \\
  V58      & ?\tablefootmark{a}     & 20 53 27.12 & -12 32 13.9 & 6166.76                   & 0.285  \\[1pt]
  \hline
 \end{tabularx}
\tablefoot{
The celestial coordinates correspond to the epoch of the reference image, which is the heliocentric Julian date $\sim$2456167~d. The epoch of maximum light is given as a heliocentric Julian date (2450000~+) in column 5 and the period is given in column 6.
\tablefoottext{a}{We are unable to classify this variable (see Sec. \ref{sec:results})}
}
\end{table}

\begin{figure}
\centering
\includegraphics[width=\linewidth]{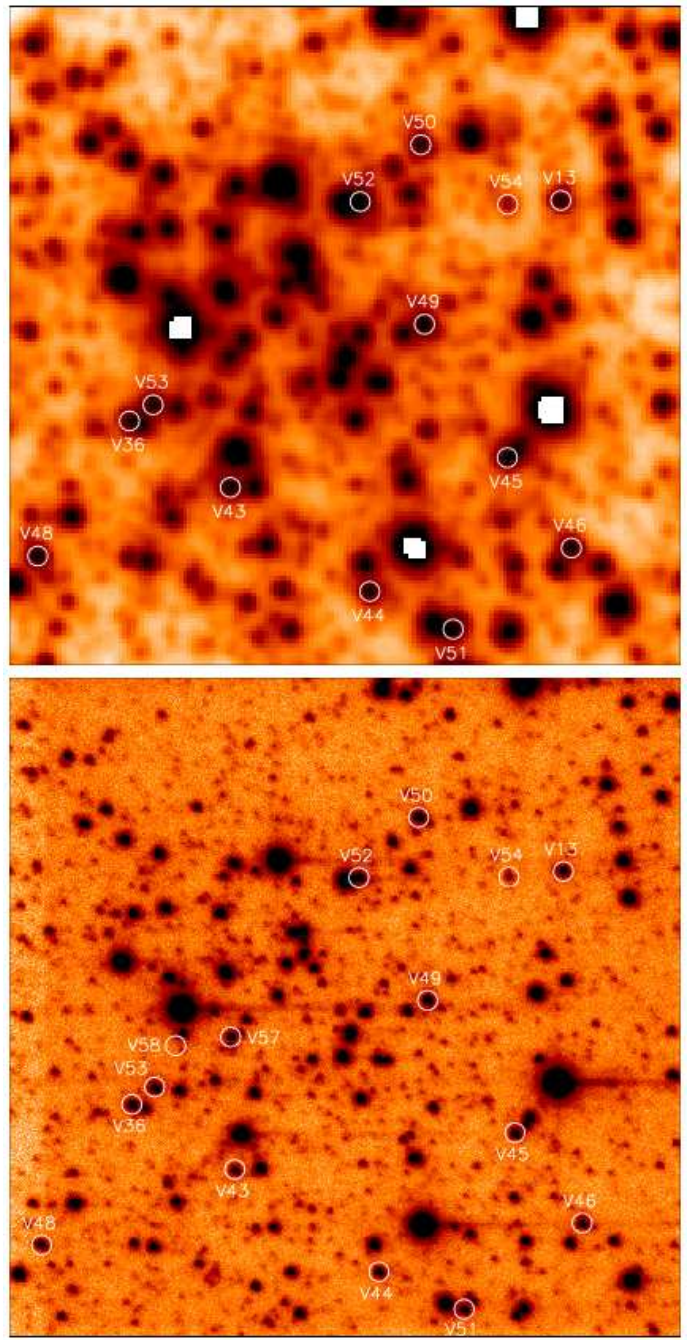}
\caption{Top: A cut-out of the B11 V reference image corresponding to the field of view of the EMCCD camera. 
         Bottom: Finding chart (constructed using our reference image) for the variables confirmed by B11 and the two new variables V57 and V58 (note that V58 is not the star that is located at the upper right edge of the circle).
         North is up and East is to the right. The image size is $45 \times 45$ arcsec$^2$. Notice the greatly improved resolution
         compared to the B11 finding chart.}
\label{fig:ref_image}
\end{figure}

In order to detect new variable stars in our data, we constructed and visually inspected an image representing the sum of the absolute-valued difference images with pixel values in units of sigma. We found two new variable stars which we assign names V57 and V58, and the details of which are given
in Table \ref{table:details}. Both stars are located close to a bright star as can be seen in Fig. \ref{fig:ref_image}. In 
the B11 data, both of these variables are within the area that cannot be measured because of the saturated pixels from the bright star. 
Using the saturation limits from B11 it can be concluded that the bright star is brighter than 14th magnitude in V. In our data we find that the bright star is about 4 magnitudes brighter than the RR Lyrae stars, which suggest that it is V$\sim13$ mag.

\begin{figure}
   \centering
   \subfloat[][V57]{\centering
                \includegraphics[width=\linewidth]{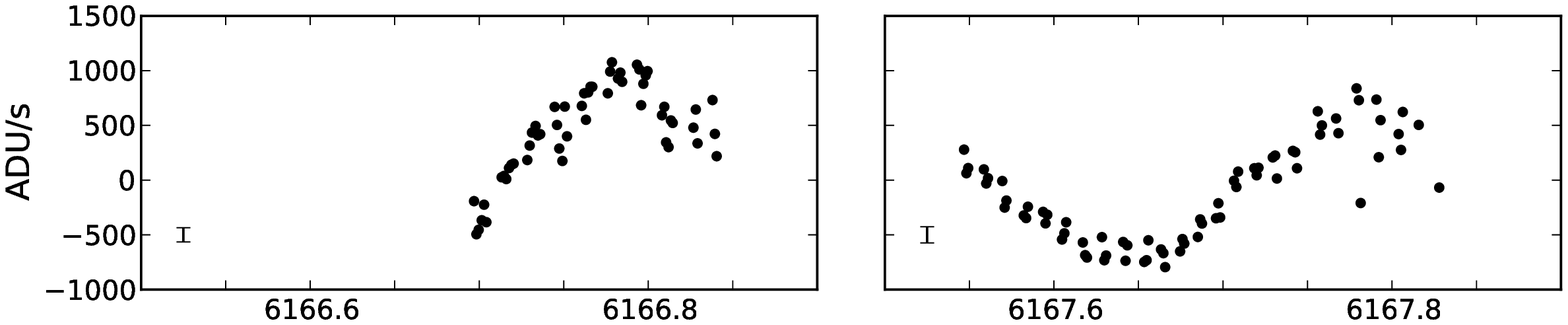}
                \label{fig:lc_v57}} \newline
   \vspace{-0.5pt}
   \subfloat[][V58]{\centering
                \includegraphics[width=\linewidth]{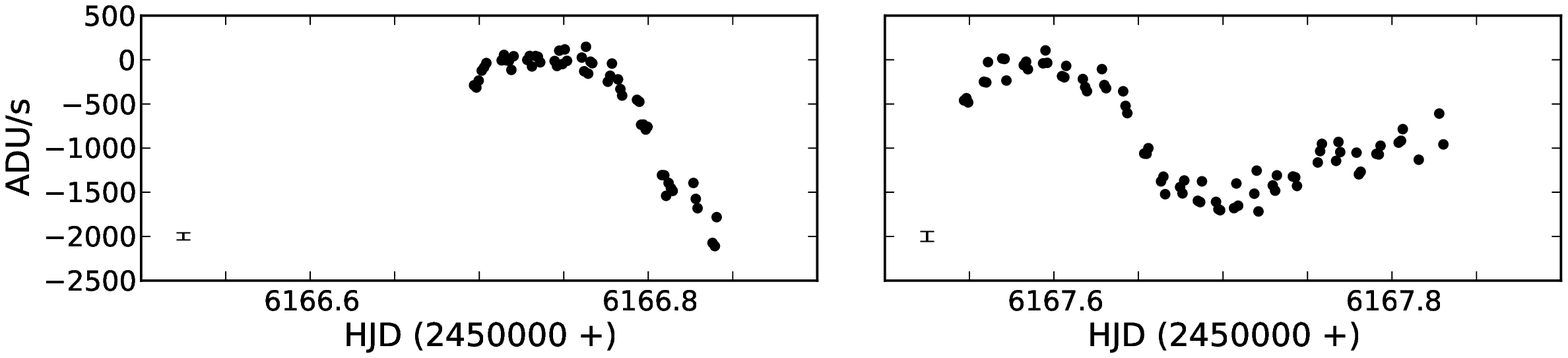}
                \label{fig:lc_v58}}
      \caption{Light curves, plotted in differential flux units, for the two new variable stars. Left and right panels show the first and second nights, respectively.
	      The typical photometric uncertainty is plotted as an error bar in each panel. 
              }
        \label{fig:lc}
   \end{figure}

\begin{figure}
   \centering
   \includegraphics[width=\linewidth]{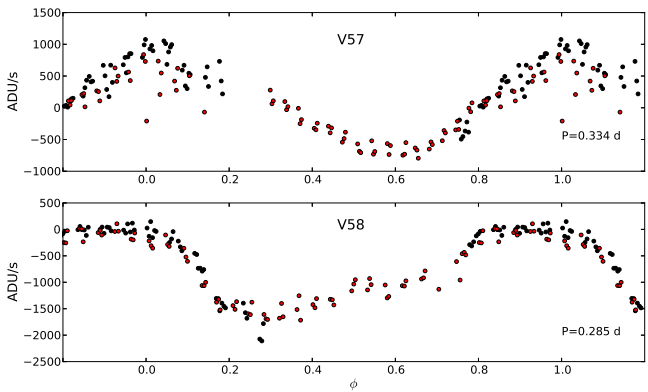}
      \caption{Phased light curves for the two new variable stars, plotted in differential flux units and using the periods from Table \ref{table:details}. 
	       Black and red dots represent the data from the first and second nights of observation, respectively. The typical uncertainty in the period is about 0.01 d for both variables
              }
        \label{fig:phased_lc}
   \end{figure}

The light curves for the two variables are shown in Fig. \ref{fig:lc}. There is increased scatter towards the end of each night which is
due to a combination of high airmass and deteriorating seeing. The variable star periods were estimated using the string-length statistic $S_Q$
\citep{Dworetsky1983} and the phased light curves are shown in Fig. \ref{fig:phased_lc}.

{\bf V57}: With a period of 0.334 days, a sinusoidal-like light curve, and a brightness on the reference image similar to that of the other RR Lyrae stars, we can safely
classify this variable as a first-overtone RR Lyrae star (RR1). 

{\bf V58}: This object has no detectable PSF-like peak in the reference image even though it shows clear PSF-like variations in the difference images. Hence the associated object
is fainter than the cluster RR Lyrae stars. The period of 0.285 days is typical of an RR1 star, although the light curve clearly deviates somewhat from being sinusoidal with relatively flat peaks and sharp drops in intensity. This star could be an eclipsing binary or an RR1 star behind the cluster. However, due to the lack of decisive evidence for either classification,
we prefer to leave the variable as unclassified.

The discovery of a new RR1 variable in NGC~6981 changes the mean period of the RR1 stars from 0.308~d (B11) to 0.312~d. The updated ratio of the number of RR1 to RR Lyrae stars is found to be
$\sim$0.17 (compared to $\sim$0.14 in B11). Both of these quantities still agree very well with the classification of NGC 6981 as an Oosterhoff type~I cluster \citep{Smith1995}.

\section{Conclusions}

Using EMCCD data with DIA we found two previously unknown variable stars in the crowded central region of the globular cluster NGC~6981. We have classified one
variable as a first-overtone RR Lyrae and we have been unable to classify the other. The discovery of the new RR1 star consolidates the classification of NGC 6981 as an
Oosterhoff type~I cluster.

Both variables are located in a crowded field and close to a much brighter star. The previous study by B11 employing conventional CCD data with DIA failed to find these variables.
Our discovery of these new variables in a carefully studied globular cluster is thus one of the first results to demonstrate the power of EMCCDs for high-precision time-series photometry in crowded fields. 
This means that EMCCDs can improve the results in a number of areas in astrophysical research, for instance the search for Earth-mass exoplanets in gravitational microlensing events, or, as mentioned here, a better constraint on the physical parameters of globular clusters.

\begin{acknowledgements}
JS acknowledges support from the ESO 2012 DGDF for a two month visit to ESO Garching.
The operation of the Danish 1.54m telescope is financed by a grant to UGJ from the Danish Natural Science Research Council. 
We also acknowledge support from the European Community's Seventh Framework Programme (FP7/2007-
2013/) under grant agreement Nos. 229517, 268421 and 274889 and from the Center of Excellence
Centre for Star and Planet Formation (StarPlan) funded by The Danish National Research
Foundation.
KAA, MD, DMB, CL, MH, RAS and CS are thankful to Qatar National Research Fund (QNRF), member of Qatar Foundation, for support by grant NPRP 09-476-1-078.
YD, AE under JS acknowledge support from 
the Communaut\'e{} fran\c caise de Belgique -
Actions de recherche concert\'ees - 
Acad\'emie universitaire to Wallonie-Europe.
TCH acknowledges support from the Korea Research Council of
Fundamental Science \& Technology (KRCF) via the KRCF Young Scientist
Research Fellowship Programme and for financial support from KASI
travel grant number 2013-9-400-00.
MR acknowledges support from FONDECYT postdoctoral fellowship No. 3120097.
\end{acknowledgements}

\bibliographystyle{aa}
\bibliography{Litt}   

\end{document}